\documentclass[twoside,14pt,a4paper]{article}

\usepackage{epsfig}
\usepackage{rotate}
\usepackage{latexsym}

\sloppypar

\renewcommand{\slash}[1]{/ \!\!\! {#1}}

\begin{document}


\begin{center}

\Large{\bf{Pion Electroproduction and Pion-induced Dileptonproduction on the Nucleon
\footnote{Work supported by BMBF, DFG and GSI Darmstadt}
}}

\large{G. Penner, T. Feuster and U. Mosel}

\textit{Institut f\"ur Theoretische Physik, Universit\"at Giessen}

\textit{D-35392 Giessen, Germany}

\vspace{1cm}

\textbf{Abstract}
\end{center}

\large
By introducing electromagnetic formfactors in the spacelike region we extend an effective, gauge invariant Lagrangian model considering Born terms, baryon resonances up to 1.7 GeV and vector meson contributions ($\rho$, $\omega$) [1] to calculate electroproduction of $\pi$-mesons. This model forms the basis for predictions of pion induced dilepton production on the nucleon. Therefore, the implemented formfactors are constructed in such a way that their analytic continuation into the timelike region is possible. It is shown that the seagull term and the $N^*(1520)$ resonance play a dominant role in the dilepton production.

\vspace{1cm}
\begin{center}
\textbf{Introduction}
\end{center}

The formfactor of the nucleon is in the focus of ongoing interest. There have been many theoretical investigations (e.g. \cite{iachello, hohler, gari, schafer, mergell, bijker}) to describe its $k^2$ behavior, which has been measured in the spacelike and timelike region for $k^2 > 4 m^2$ with the nucleon mass $m$. Here, we examine the possibility to determine the nucleon formfactor in the ``forbidden'' timelike region through a half-offshell reaction: the pion induced dilepton production, which will be investigated experimentally at GSI Darmstadt with the new dilepton spectrometer HADES. The understanding of this reaction is necessary for making reliable predictions for pion induced reactions on nuclei \cite{weidmann}. In these reactions, the dileptons will be used as microscopic probes for investigating in-medium modifications of vector meson masses.
\newline
To be free of ambiguities in the model all relevant parameters have to be fixed using other reactions. Therefore, we extend a model developed to investigate photoproduction of pions on the nucleon \cite{feuster} to a description of virtual photons. The additional electromagnetic coupling constants and formfactor parameters are determined by examining pion electroproduction and available formfactor data. Due to the structure of both reactions, it is possible to separate them into a leptonic and a hadronic part. The hadronic parts of these two reactions are easily related to each other by detailed balance. This is shown in Fig.\ \ref{picintro}. Thus, we are able to use exactly the same hadronic tensor to describe pion induced reactions including virtual photons as in pion electroproduction. 

\vspace{1cm}
\begin{center}
\textbf{Review of the Model}
\end{center}

The model, which is described in detail in \cite{feuster}, contains nucleon ($N$) and nucleon resonance ($R$) s-channel and u-channel graphs, $\rho$, $\omega$ and $\pi$ t-channel graphs and the seagull graph. For the nucleon and the spin-$\frac{1}{2}$ resonances ($N^*(1440)$, $N^*(1535)$, $\Delta^*(1620)$ and $N^*(1650)$) we use pseudovector pion coupling and for the spin-$\frac{3}{2}$ resonances the usual coupling, incorporating offshell projectors. 
\newline
For the case of virtual photons one has to extend the existing model, since there are additional couplings that do not contribute for real photons, i.e. that are proportional to $k^2$. Furthermore, we apply the Gross-Riska procedure \cite{gross}  to the Born terms to ensure gauge invariance even after having introduced formfactors. Otherwise, one would get unphysical restrictions on the different formfactors, i.e. the same $k^2$ dependence for the pion, seagull and the $F_1^V$ formfactor (cf. \cite{nozawa}). This means to replace in the nucleon electromagnetic coupling
\begin{eqnarray}
\Gamma^\mu_{NN \gamma}(k^2) &=& - \mbox i e \left( F_1^N (k^2) 
	\gamma^\mu - F_2^N (k^2) \frac{\sigma^{\mu \nu} 
	\partial_\nu}{2 m} + \mbox i F_3^N (k^2) \partial^\mu \right) 
	\nonumber
\end{eqnarray}
the formfactor $F_3^N$ by $(F_1(0)^N - F_1^N(k^2)) \slash \partial / \partial^2$ and similarly for $F_2^\pi$ and $F_2^S$. Here the derivative $\partial$ acts on the electromagnetic field $A_\mu$. 
\newline
All the relevant hadronic coupling constants and the electromagnetic ones contributing to photoproduction have been taken from fit \# 5 in \cite{feuster}. 

\vspace{1cm}
\begin{center}
\textbf{Electromagnetic Formfactors}
\end{center}

To be able to describe space- and timelike photons we use a simple, VMD motivated model for all relevant electromagnetic formfactors. Furthermore, we implement $\rho$-$\omega$-mixing, which arose in the analysis of the pion formfactor from the need to describe the small kink at the $\omega$ mass in the otherwise smooth data. 
\newline
Hence the pion formfactor reads
\begin{eqnarray}
F_\pi (k^2) = F_\pi (0) + \frac{g_{\rho \pi}}{g_{\rho \gamma}} \frac{k^2}
	{m_\rho^2 - k^2 - \mbox i m_\rho \Gamma_\rho (k^2)} + 
	\frac{\epsilon g_{\rho \pi}}
	{g_{\omega \gamma}} \frac{k^2}
	{m_\omega^2 - k^2 - \mbox i m_\omega \Gamma_\omega}
	\nonumber \; ,
\end{eqnarray}
where the complex mixing parameter $\epsilon$ is small: $\epsilon = \left| \epsilon \right| \mbox e^{\mbox {\scriptsize i} \varphi}$ with $\left| \epsilon \right| \approx 0.0315$ and $\phi \approx 100.5^\circ$. For further details on $\rho$-$\omega$-mixing see e.g. \cite{oconnell}. The energy dependence of the $\rho$ width is parameterized through the 99\% dominating two-pion-decay, 
while the $\omega$ width is assumed to be constant due to its smallness of $\Gamma_\omega = 8.43$ MeV. 
\newline
This simple scheme is not sufficient for describing the dipole behavior of the baryon formfactors, i.e. the nucleon formfactors. In this case we assume a two step process, so that the gamma not only couples via vector meson dominance, but as well to an extended core which is described by an intrinsic formfactor $F_{cut}(k^2)$. There are several works using similar formfactor models for the nucleon \cite{iachello, gari, bijker, williams} and some more sophisticated models (e.g. \cite{mergell}) as well, but we are aiming for a description which is easily extendable to the timelike region and the baryon resonances. Hence, we use a simple shape for the intrinsic formfactor:
\begin{eqnarray}
F_{cut} (k^2) = \frac{\Lambda^4 + \frac{1}{4} \tilde m_V^4} 
	{\Lambda^4 + (k^2 - \frac{1}{2} \tilde m_V^2)^2} 
	\qquad \mbox{with} \qquad 
	\tilde m_V = \frac{1}{2} (m_\rho + m_\omega) \qquad .
	\nonumber
\end{eqnarray}
This has an asymptotic dipole behavior for $k^2 < 0$ and is polefree for all $k^2$. Furthermore, the coupling to a real photon is reproduced for $k^2 = 0$ (i.e. $F_{cut} (0) = 1$) and for $k^2 = \tilde m_V^2$ the coupling corresponds to the one to a real vector meson ($\rho$, $\omega$): $F_{cut} (\tilde m_V^2) = 1$. Now the nucleon formfactors is given by
\begin{eqnarray}
F_1^V (k^2) &=& {F_{cut}}_1 (k^2) \left( \frac{1}{2} + \frac{g_{NN \rho}}
	{g_{\rho \gamma}} \frac{k^2}{m_\rho^2 - k^2 - \mbox i m_\rho 
	\Gamma_\rho (k^2)} \right.	\nonumber \\
&& \hspace{4cm} \left. + \frac{\epsilon g_{NN \rho}}
	{g_{\omega \gamma}} \frac{k^2}{m_\omega^2 - k^2 - \mbox i m_\omega 
	\Gamma_\omega} \right)		\label{nucform} 
\end{eqnarray}
and in the same way for $F_1^S (k^2)$, $F_2^V (k^2)$ and $F_2^S (k^2)$. The resulting shape is displayed in Fig. \ref{picnucform}.
\newline
Since for none of the resonances implemented here a $N \omega$ decay has been observed, we assume that all resonances couple to an $\omega$ meson only through $\rho \omega$ mixing. The necessary coupling constants for the $R N \rho$ decay are calculated with the help of the measured decay rates \cite{manley}. To avoid changing the phase between the different couplings in comparison with the photoproduction amplitudes and to reproduce the vector meson coupling for $k^2 = \tilde m_V^2$, the electromagnetic formfactors of the nucleon resonances are
\begin{eqnarray}
F_{elmg} (k^2) = && \hspace{-7mm} F_{cut} \left( 1 + 
	\frac{g_{R N \rho}}{g_{\rho \gamma} g_{R N \gamma_i}} 
	\frac{k^2}{m_\rho^2 - k^2 - \mbox i m_\rho 
	\Gamma_\rho (k^2)} \right.	\nonumber \\
&& \hspace{4cm} \left. + \frac{\epsilon g_{R N \rho}}
	{g_{\omega \gamma} g_{R N \gamma_i}} 
	\frac{k^2}{m_\omega^2 - k^2 - \mbox i m_\omega 
	\Gamma_\omega} \right)	\; .	\nonumber
\end{eqnarray}
This is in agreement with quark counting rules and an analysis by Stoler \cite{stoler} which shows that the formfactors of all resonances in the spacelike sector are remarkably similar.
\newline
The remaining undetermined parameters -- the cutoff parameter and the electromagnetic resonance coupling to virtual photons -- are fixed by two different schemes: In the $\Delta (1232)$ case we use the relation between the Sachs decomposition of the formfactors and the one used here, whereas for all other resonances we compare with data for helicity amplitudes.
\newline
Finally, there are still two formfactors to be set: the axial formfactor of the Kroll-Rudermann graph and the formfactor of the $\rho / \omega \pi \gamma$ vertex. Since the axial coupling resembles -- besides a parity operator $\gamma_5$ -- the $F_1$ nucleon coupling, the axial formfactor is taken to be:
\begin{eqnarray}
F_A^V (k^2) &=& {F_{cut}} (k^2) \left( \frac{1}{2} + \frac{g_{NN \rho}}
	{g_{\rho \gamma}} \frac{k^2}{m_\rho^2 - k^2 - \mbox i m_\rho 
	\Gamma_\rho (k^2)} \right.	\nonumber \\
&& \hspace{4cm} \left. + \frac{\epsilon g_{NN \rho}}
	{g_{\omega \gamma}} \frac{k^2}{m_\omega^2 - k^2 - \mbox i m_\omega 
	\Gamma_\omega} \right)		\nonumber
\end{eqnarray}
and in the same way for $F_A^S (k^2)$. The cutoff parameter $\Lambda_1$ has the same value as in the nucleon case. For the vector meson contributions in the $t$-channel again the concept of VMD is adopted. The necessary coupling constant can be extracted either with the three-pion decay of the $\omega$ \cite{goldberg} or with superconvergent sum rules, either way gives a value of $g_{\omega \pi \rho} \approx 3$. From quark counting rules one would expect a monopole falloff of the $\omega \pi \gamma$ formfactor, but experimental results indicate a different behavior \cite{landsberg}. We nevertheless retain the VMD concept and use an additional cutoff function in the same way as for the nucleon. However, since the contribution of the $t$ channel $\omega$ graph on the vector meson masses is small compared to the nucleon contribution, the results do not depend on the choice of this specific formfactor. 

\vspace{1cm}
\begin{center}
\textbf{Pion Electroproduction}
\end{center}

In order to fix the relevant transition matrix element we first consider the electroproduction of pions. Following \cite{nozawa}, one can rewrite the differential cross section of the pion electroproduction by using gauge invariance and finds:
\begin{eqnarray}
\frac{\mbox d ^3 \sigma}{\mbox d {E'_e}^* \mbox d {\Omega'_e}^* \mbox d 
	\Omega_\pi} 
	= \Gamma_t \left[ 
	\frac {\mbox d \sigma_T}{\mbox d \Omega_\pi} + 
	\varepsilon \frac {\mbox d \sigma_L}{\mbox d \Omega_\pi} +
	\varepsilon \frac {\mbox d \sigma_P}{\mbox d \Omega_\pi} 
	\cos 2 \varphi_\pi + 
	\sqrt {2 \varepsilon ( 1 + \varepsilon )} 
	\frac {\mbox d \sigma_I}{\mbox d \Omega_\pi} \cos \varphi_\pi
	\right] \qquad , \nonumber 
\end{eqnarray}
where $\Gamma_t$ is of electromagnetic origin and $\varepsilon$ is the polarization of the virtual photon. The various cross sections are denoted by transversal, longitudinal, polarization and interference cross section, depending on which parts of the hadronic tensor are contained. 
\newline
We start off with calculations in the $\Delta (1232)$ region, where it is possible to compare with other models (e.g. \cite{nozawa}). In this region the main contributions arise from the $\Delta (1232)$ resonance and the background caused by the Born graphs (and the vector meson $t$-channel graphs). While \cite{nozawa} explicitly includes final-state interaction effects by half-off-energy-shell $\pi N$-scattering matrix elements, these are here just included by using effective coupling constants and imaginary contributions in the resonance propagators. As in \cite{nozawa}, we first concentrate on inclusive $p(e,e')$ cross sections. In Fig. \ref{picdelk2} the influence of the different parametrizations for the electromagnetic formfactors is displayed. Obviously, the results depend on the shape of the $\Delta (1232)$-formfactor, while in this energy region the influence of the higher resonances is still small. With the VMD formfactor the $k^2$ dependence is in good agreement with the data. The few data for the longitudinal cross section are as well reproduced.
\newline
For the $\pi^0$-production we look at two off-plane kinematics $\varphi_\pi = 45^\circ$ and $\varphi_\pi = 90^\circ$, at which the interference and the polarization cross sections, respectively, become important. The results (cf. Fig. \ref{picdel5al}) agree with respect to the maximum height and position with the data. From this we can conclude that within the $\Delta (1232)$ region our model is able to describe the dynamics of the electroproduction of pions in agreement with experiment.
\newline
To be able to make reliable predictions for the dilepton production in the interesting vector meson region, we need to proceed to center of mass energies of at least $1.7$ GeV, since the maximum invariant dilepton mass then is in the order of $m_{e^+ e^-} \approx 0.780$ GeV. In this energy region all the other resonance contributions to pion electroproduction become more and more important and the $\Delta (1232)$ loses its influence. This can be very easily seen from Fig. \ref{picgdt18} where the total $p(e,e'\pi^0)p$ cross section is displayed. The model does not correctly reproduce the shift of the $\Delta (1232)$ peak, while the shape and magnitude for higher energies are correctly described. 
\newline
Finally, we get to the most interesting electroproduction reaction, the $\pi^-$-production. Since the experiments planned at GSI will use a $\pi^-$-beam for producing dileptons, the $e n \rightarrow e' p \pi^-$-reaction is the one to be focussed on. Unfortunately, there exist only few data for this reaction: Usually, only the ratio between $e n \rightarrow e' p \pi^-$ and $e p \rightarrow e' n \pi^+$ are measured with the help of deuteron targets. The results for the differential cross sections in dependence of the energy are shown in Fig. \ref{picnpwdif}. Obviously, the agreement gets worse for higher energies so that the energy limitation of the model becomes evident. 

\vspace{1cm}
\begin{center}
\textbf{Pion-induced Dilepton Production}
\end{center}

After having compared the electroproduction results of the model with the available data, we can now proceed to calculate the pion induced dilepton production. The focus is on the $\pi^-$ induced channel, as explained above. In the following, we just look at an CMS energy of $1.75$ GeV. The most important contributions of all diagrams arise from the $N^* (1520)$ resonance and the seagull graph, which is evident from Fig. \ref{picdilpnbor152}. Due to the large $\rho$ coupling of the $N^* (1520)$ one gets a broad $\rho$ shoulder. This effect is very sensitive to the coupling to virtual photons of the neutral resonance appearing in the $s$-channel, which has been determined in the electroproduction. Furtheron, the small kink on the $\omega$ mass is not - as one would expect - due to the nucleon, but to the axial formfactor. If we switch off the various contributing formfactors of the Born graphs we see that the axial formfactor clearly dominates. 
\newline
The reason for this behavior is that the $s$- and $u$-channel contributions of the nucleon almost cancel each other in this threshold scenario ($m + m_\omega \approx 1.72$ GeV with a CMS energy of $1.75$ GeV), so that the remaining cross section is dominated by the strong $\omega$ coupling entering via the seagull graph. 

\vspace{1cm}
\begin{center}
\textbf{Conclusions}
\end{center}

Employing a model already used for pion photoproduction, we have investigated its continuation to virtual, space- and timelike photons. The additionally entering formfactor parameters and coupling constants have been fixed with the help of electroproduction data and we find a reasonable agreement with the experimental data. Consequently, we have performed parameterfree calculations on the dilepton production on the nucleon with special emphasis on the $\pi^-$ induced reaction. We find a strong domination in this channel of two graphs: the $s$-channel of the $N^* (1520)$ resonance and the Kroll-Rudermann term required by gauge invariance for pseudovector coupling at the $NN\pi$-vertex. 

\vspace{1cm}
\begin{center}
\textbf{References}
\end{center}

\def\refname{}

\vspace{1cm}
\begin{center}
\textbf{Figures}
\end{center}


%
\begin{figure}[hbt]
\caption[]
{\label{picintro} \parbox[t]{13cm}
{Illustration of pion electroproduction (\textit{left}) and pion-induced dilepton production (\textit{right}).}}
\end{figure}
\begin{figure}[hbt]
\caption[]
{\label{picnucform} \parbox[t]{13cm}
{Sachs formfactors of the nucleon. \textit{Upper left:} proton magnetic formfactor. \textit{Upper right:} proton electric formfactor. \textit{Down left:} neutron magnetic formfactor. \textit{Down right:} neutron electric formfactor. $\cdots$ represents the dipolfit, --- the form calculated with (\ref{nucform}).}}
\end{figure}
\begin{figure}[hbt]
\caption[] 
{\label{picdelk2} \parbox[t]{13cm}
{$k^2$ dependence of the transversal inclusive cross section in the $\Delta(1232)$ region. \textit{Left:} for a CMS energy $\sqrt s = 1.22$ GeV, \textit{right:} for $\sqrt s = 1.27$ GeV. $\cdots$: calculation from \cite{nozawa}, $\mbox - \mbox - \mbox -$: calculation with the formfactors used in \cite{nozawa}, $\mbox - \cdot \mbox - \cdot$: calculation with VMD formfactors, ---: calculation with all resonances and $\mbox - \cdot \cdot$: calculation for the longitudinal cross section considering all resonances. The data are from \cite{batzner}; the lower datapoints are measurements of the longitudinal cross section.}}
\end{figure}
\begin{figure}[hbt]
\caption[] 
{\label{picdel5al} \parbox[t]{13cm}
{The $\vartheta_\pi$ differential cross section of $e p \rightarrow e' p' \pi^0$ for fixed energy and fixed reaction plane angle $\varphi_\pi$. Illustrated is the calculation for $k^2 = 
-0.55$ GeV$^2$ and $\varepsilon = 0.965$. ---: VMD calculation with all resonances, -$\cdot$-: $\Delta (1232)$ only. The data are from \cite{lath79}. We find similar agreement for $k^2 = -0.3$ GeV$^2$ and $k^2 = -0.7$ GeV$^2$.}}
\end{figure}
\begin{figure}[hbt]
\caption[] 
{\label{picgdt18} \parbox[t]{13cm}
{Total cross section of $e p \rightarrow e' p' \pi^0$ in dependence of the CMS energy for various $k^2$. $\varepsilon \approx 0.9$ in all cases. The data are from: $\circ$: \cite{siddle}, \raisebox{0.5mm}{\tiny $\bigtriangleup$}: \cite{shuttle}, $\bullet$: \cite{lath79}, $\diamond$: \cite{alder} and \raisebox{0.5mm}{\tiny $\bigtriangledown$}: \cite{lath81}.}}
\end{figure}
\begin{figure}[hbt]
\caption[] 
{\label{picnpwdif} \parbox[t]{13cm}
{The differential cross section $\mbox d \sigma / \mbox d \Omega_\pi$ of $e n \rightarrow e' p \pi^-$ for various reaction angles $\vartheta_\pi$ in dependence of the CMS energy $\sqrt s$. In all cases we have $\varphi_\pi \approx 90^\circ$. The data are from \cite{evang} combined with \cite{morris78} ($\diamond$), from \cite{evang} with \cite{wright} ($\circ$) or from \cite{aldpn} with \cite{wright} ($\bullet$).}}
\end{figure}
\begin{figure}[hbt]
\caption[] 
{\label{picdilpnbor152} \parbox[t]{13cm}
{\textit{Left:} influence of the $N^* (1520)$ resonance on the total $\pi^- p \rightarrow n e^+ e^-$ cross section at $\sqrt s = 1.75$ GeV. ---: Calculation with the coupling constant $g_{N^* (1520)^n N \gamma^*}$ extracted from electroproduction, $\mbox - \mbox - \mbox -$: calculation without $N^* (1520)$ and $\cdots$: calculation with switched sign of $g_{N^* (1520)^n N \gamma^*}$. \textit{Right:} Influence of the Born terms on the cross section of the $\pi^-$ induced dilepton production on the proton at $\sqrt s = 1.75$ GeV. ---: Full calculation, $\mbox - \mbox - \mbox -$: calculation without Born terms, $\cdots$: calculation without nucleon formfactor and $\mbox - \cdot \mbox -$: calculation without seagull formfactor.}}
\end{figure}
%


\setcounter{figure}{0}

\begin{figure}[hbt]
\vspace{2mm}
\begin{center}
\parbox{6cm} {\epsfig{file=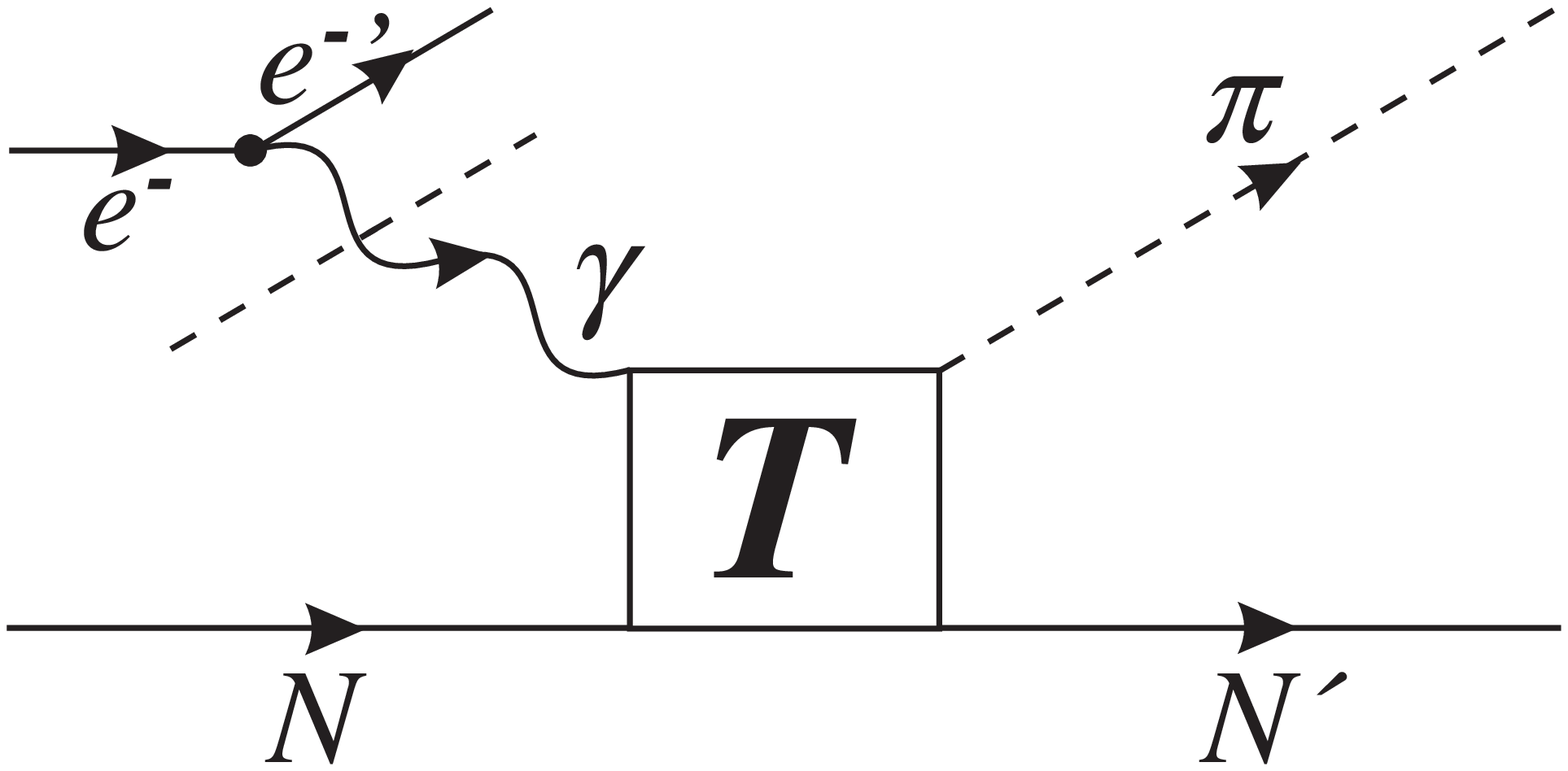, width=5cm} }
\hspace{2cm} 
\parbox{6cm} {\epsfig{file=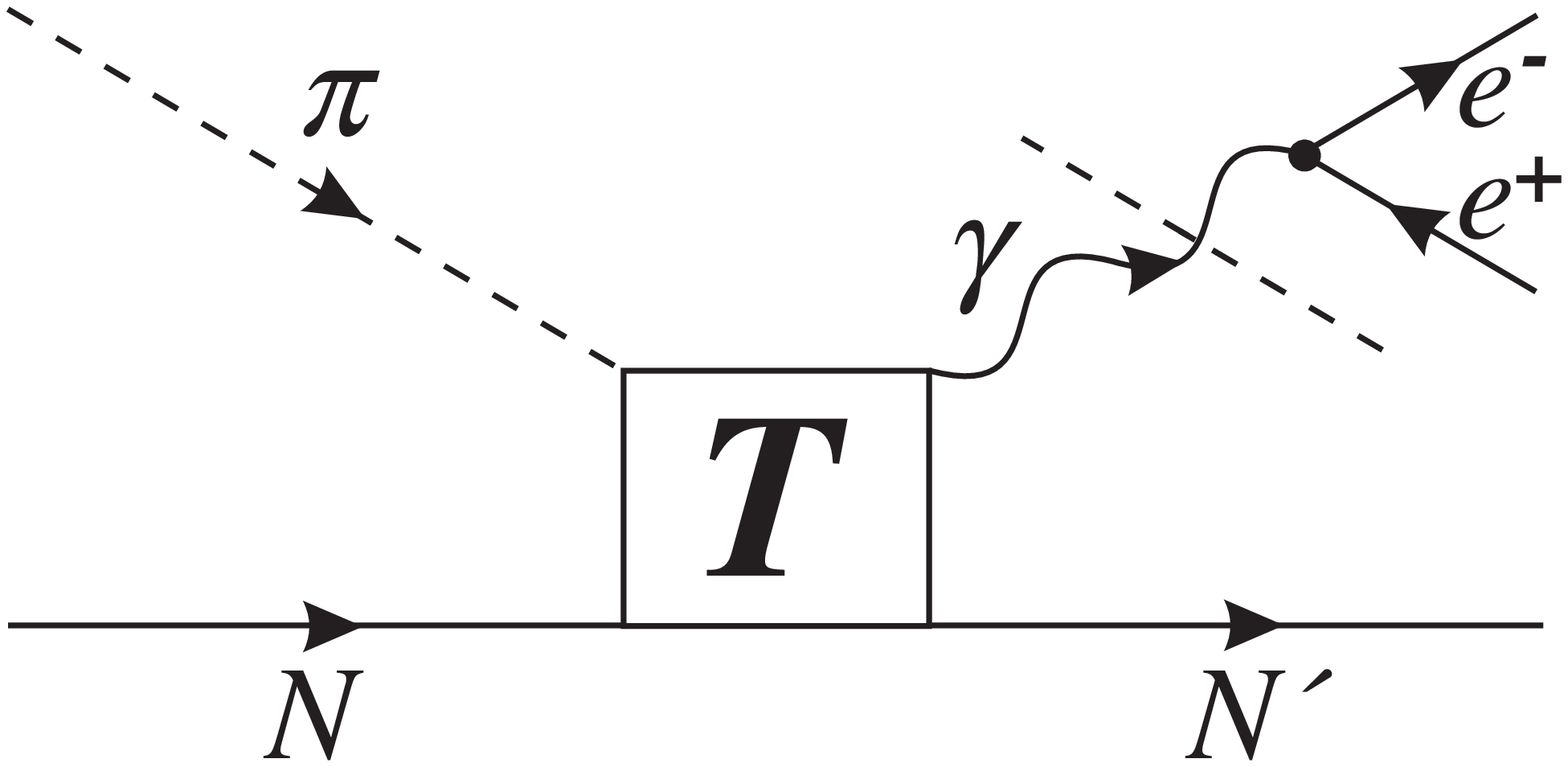, width=5cm}}
\end{center}
\vspace{-6mm}
\caption[]
{\label{picintro}}
\end{figure}
\begin{figure}[hbt]
\vspace{-5mm}
\begin{center}
\parbox{11cm}{\rotate[r]{\epsfig{file=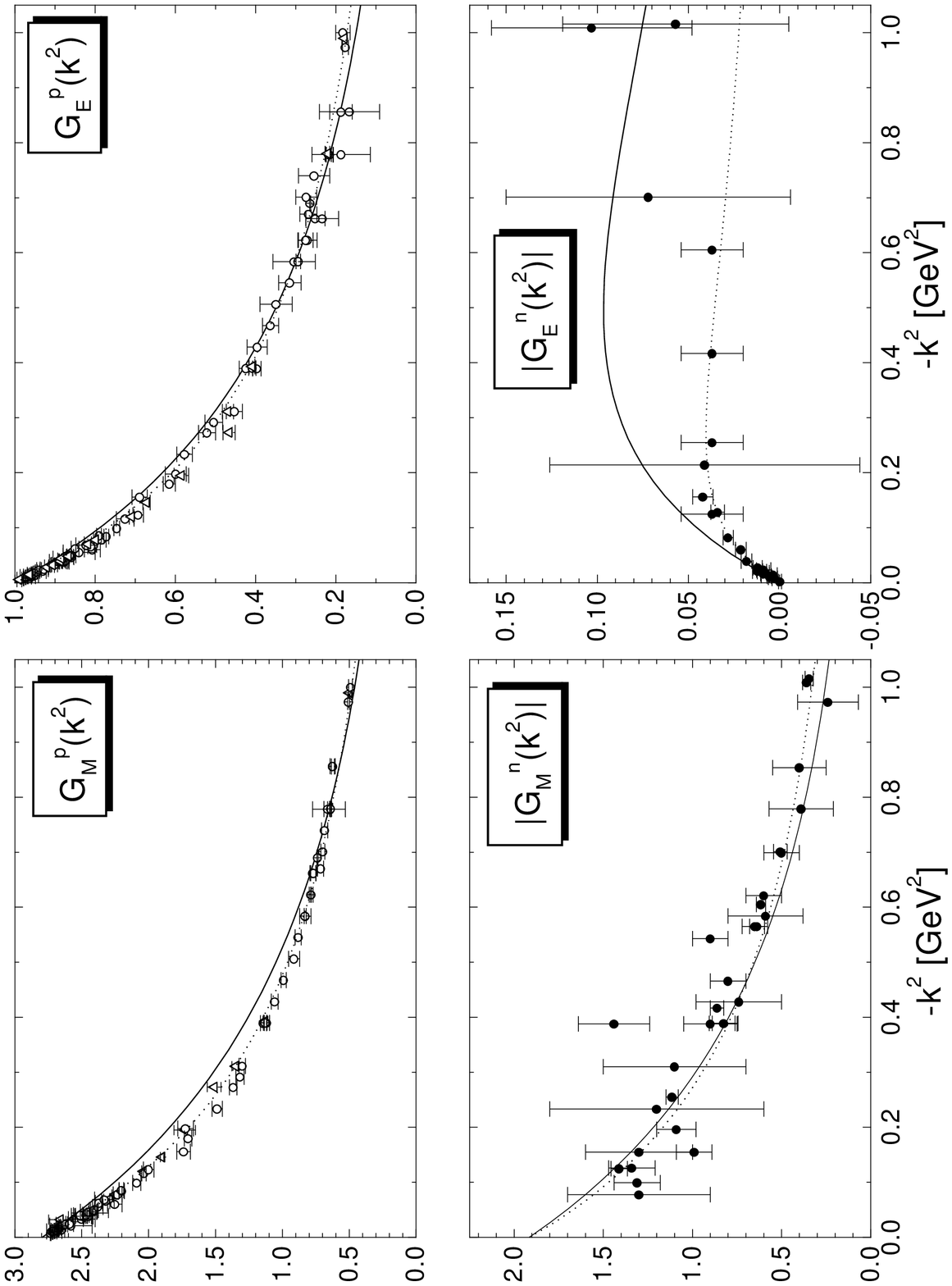, width=11cm}}}
\end{center}
\vspace{-8mm}
\caption[] 
{\label{picnucform}}
\end{figure}
\begin{figure}[hbt]
\vspace{-5mm}
\begin{center}
\parbox{8cm}{\rotate[r]{\epsfig{file=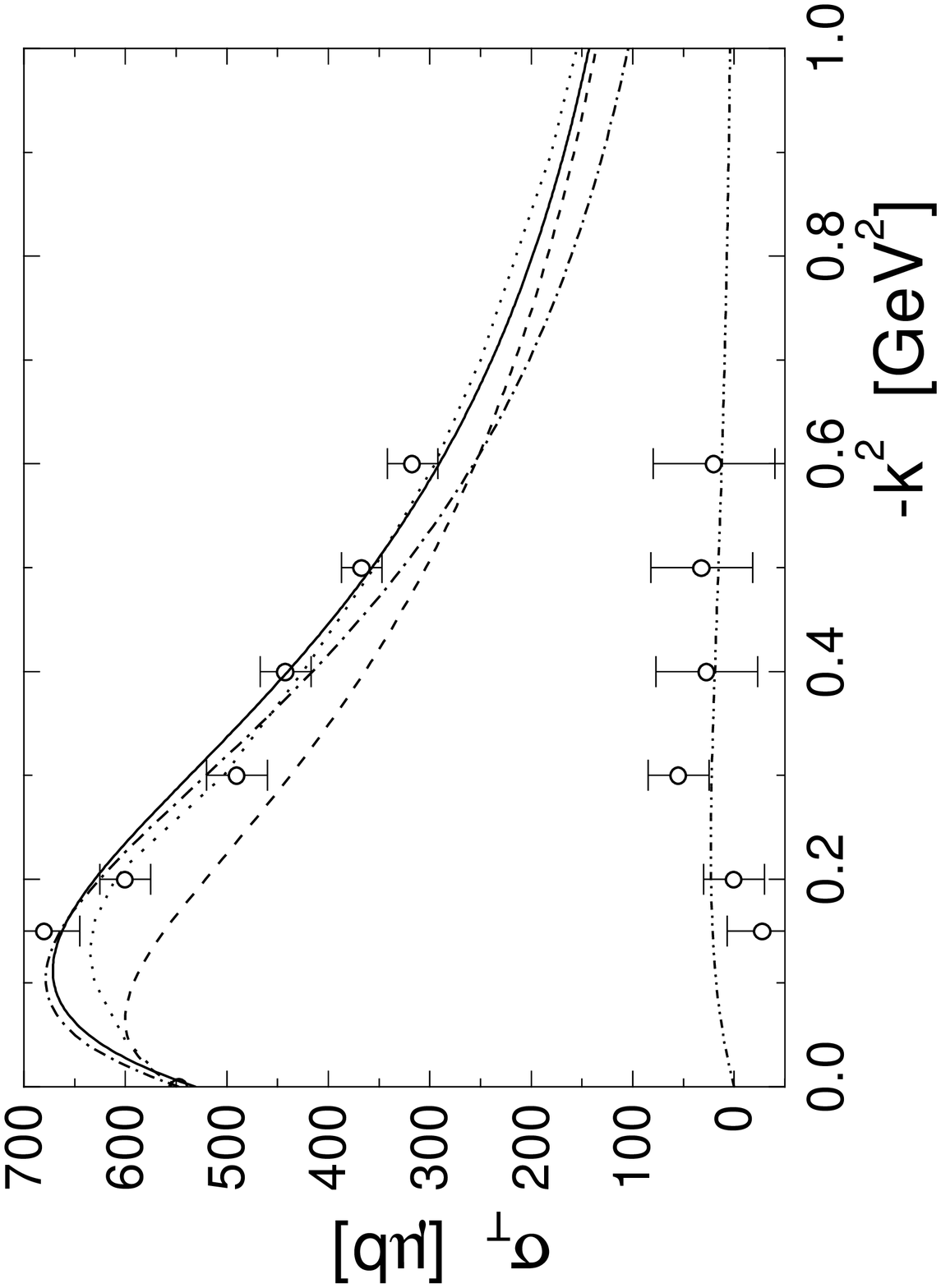, height=8cm}}}
\hspace{-2.1cm}
\parbox{8cm}{\rotate[r]{\epsfig{file=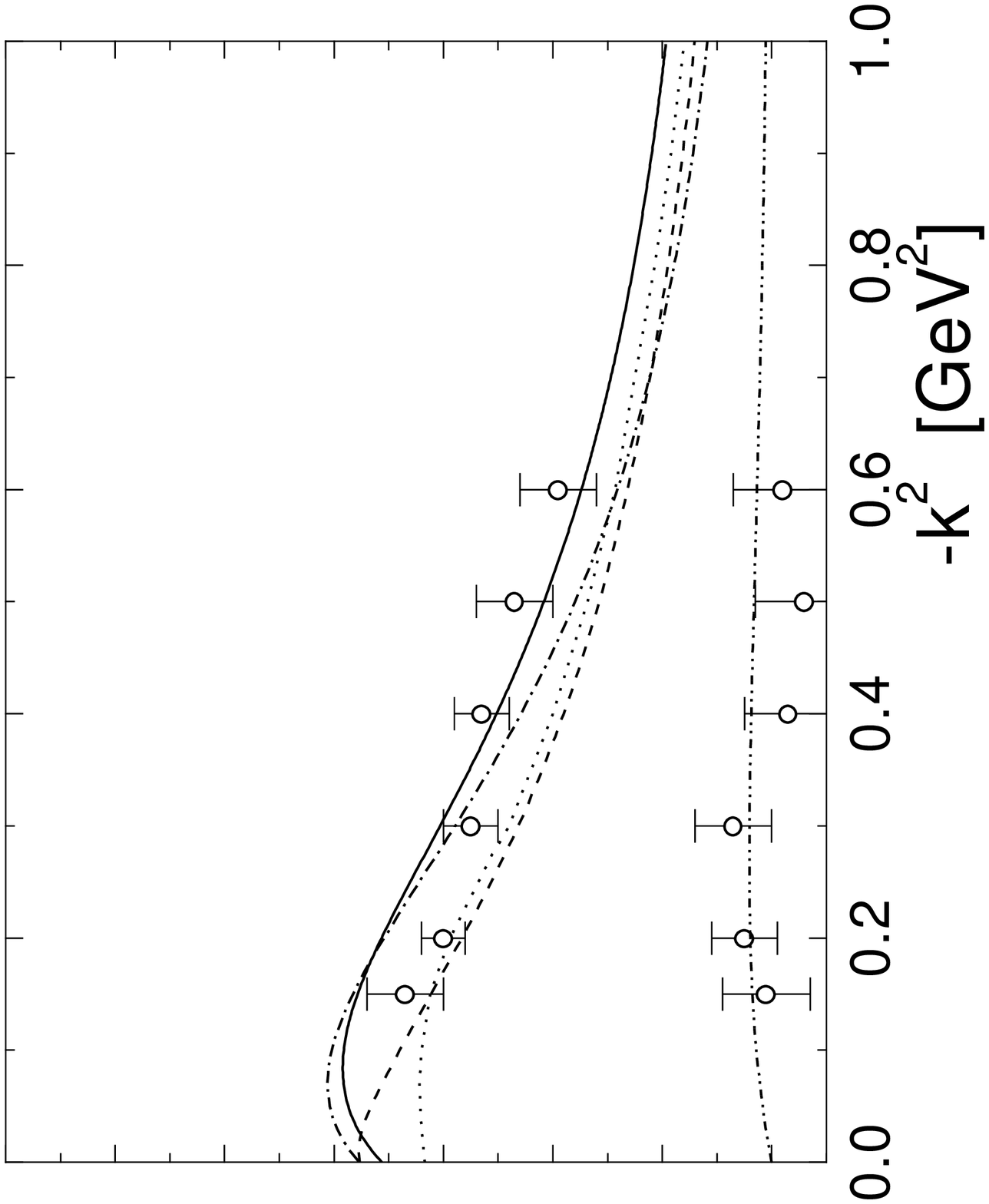, height=8cm}}}
\end{center}
\vspace{-8mm}
\caption[] 
{\label{picdelk2}}
\end{figure}
\begin{figure}[hbt]
\vspace{-1.5cm}
\begin{center}
\parbox{16cm}{\rotate[r]{\epsfig{file=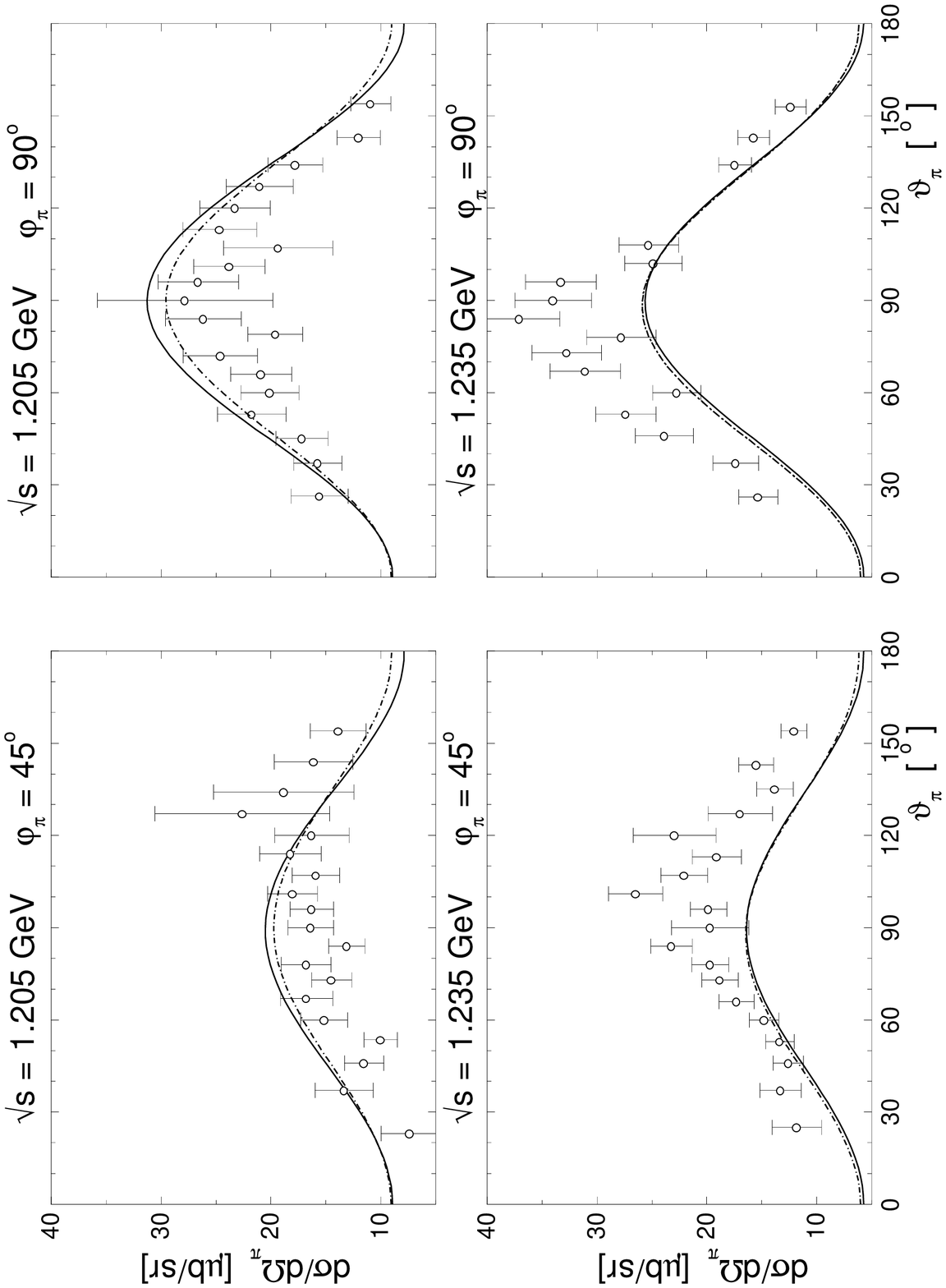, height=16cm}}}
\end{center}
\vspace{-8mm}
\caption[] 
{\label{picdel5al}}
\end{figure}
\begin{figure}[hbt]
\vspace{-1.5cm}
\begin{center}
\parbox{16cm}{\rotate[r]{\epsfig{file=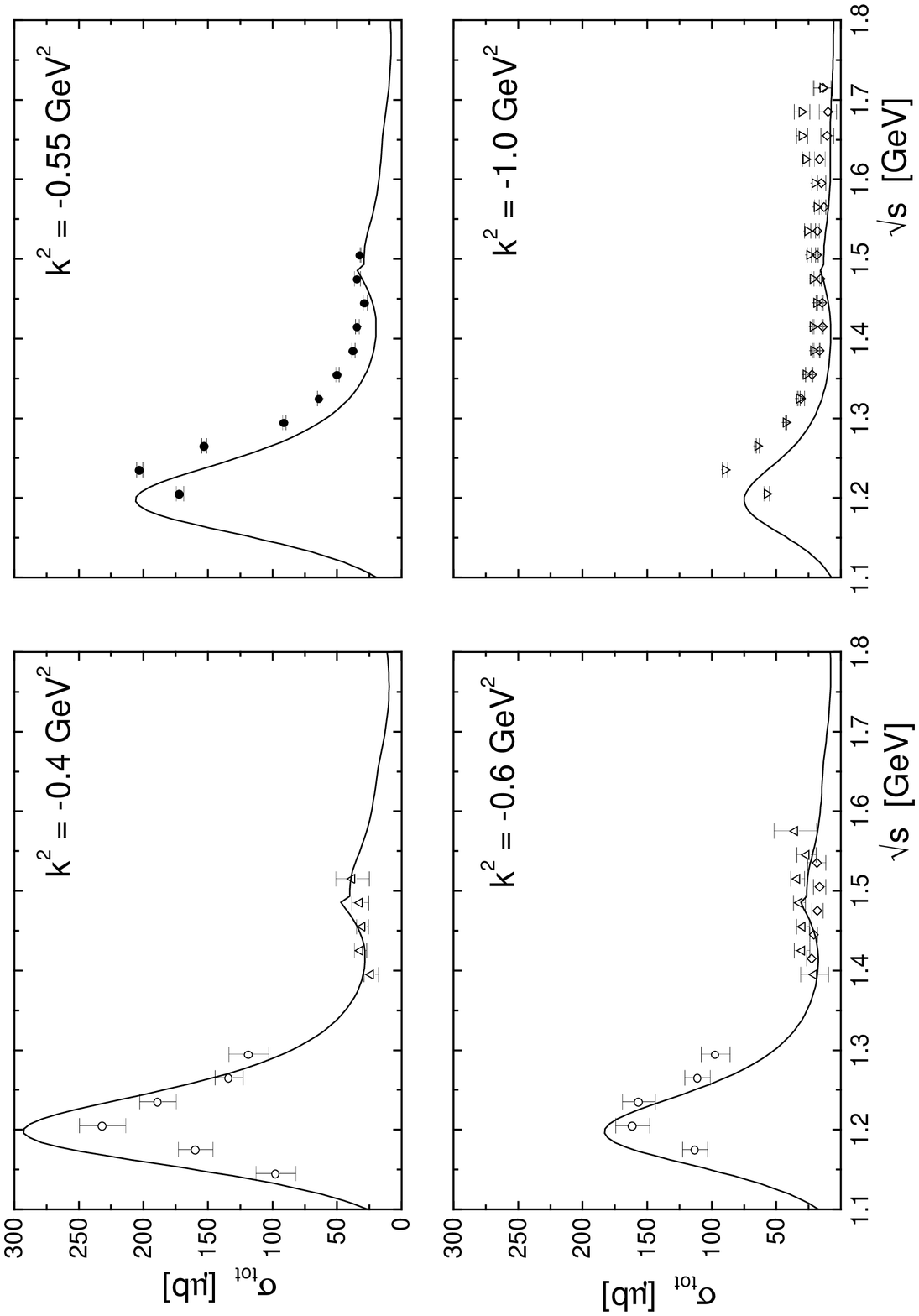, height=16cm}}}
\end{center}
\vspace{-8mm}
\caption[] 
{\label{picgdt18}}
\end{figure}
\begin{figure}[hbt]
\vspace{-5mm}
\begin{center}
\parbox{16cm}
{\rotate[r]{\epsfig{file=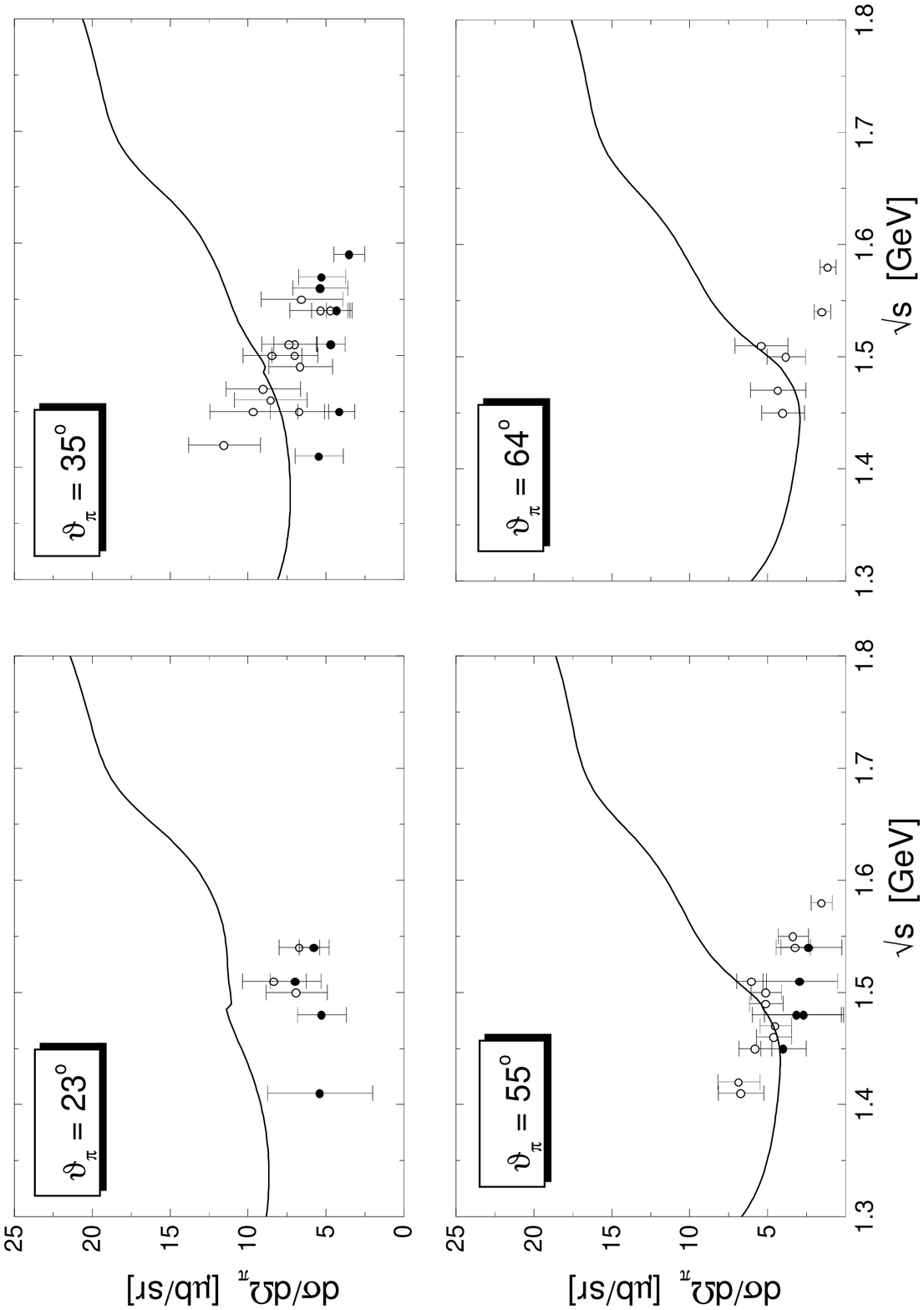, height=16cm}}}
\end{center}
\vspace{-8mm}
\caption[] 
{\label{picnpwdif}}
\end{figure}
\begin{figure}[hbt]
\vspace{-5mm}
\begin{center}
\parbox{8cm}{\rotate[r]{\epsfig{file=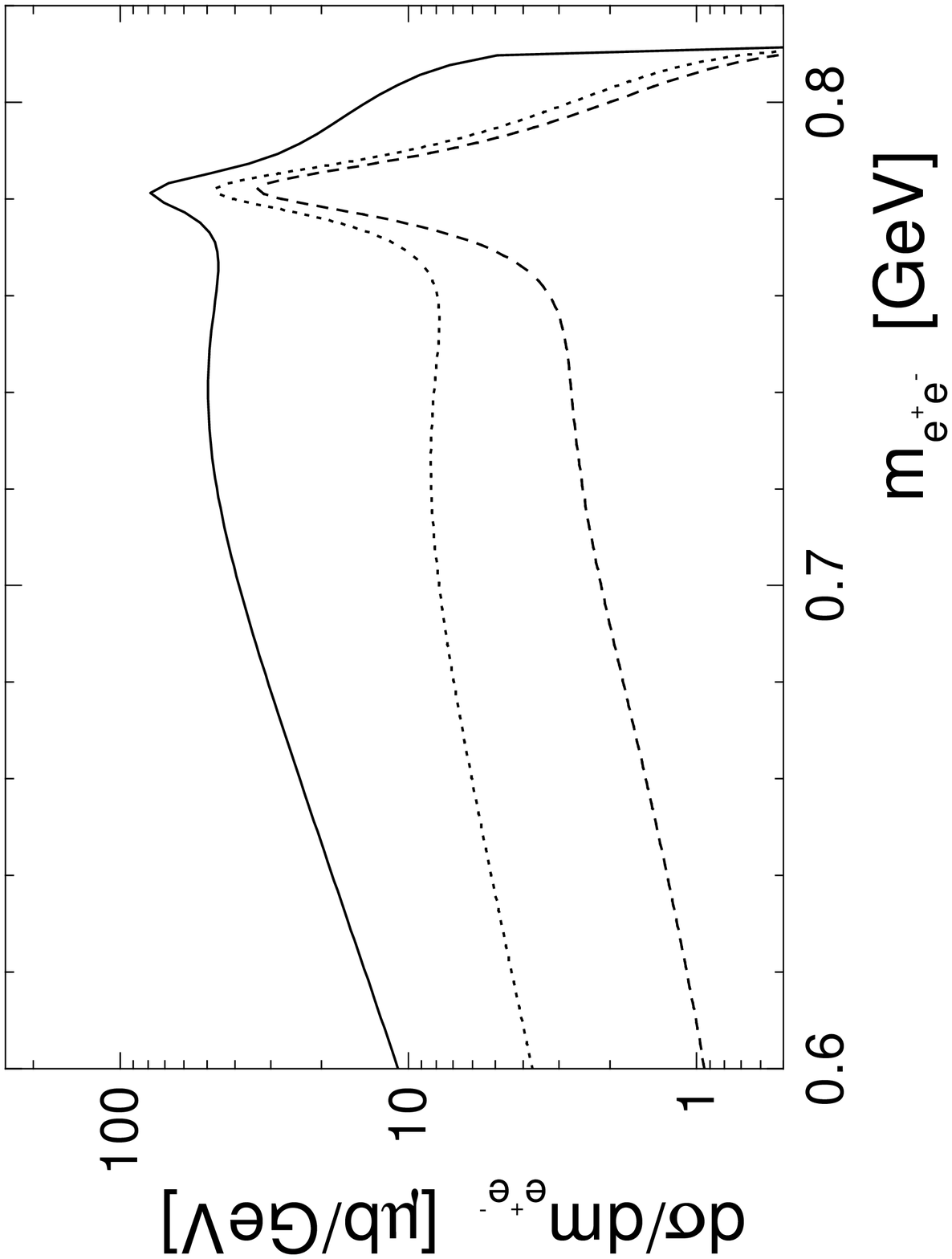, height=8cm}}}
\hspace{-2.1cm}
\parbox{8cm}{\rotate[r]{\epsfig{file=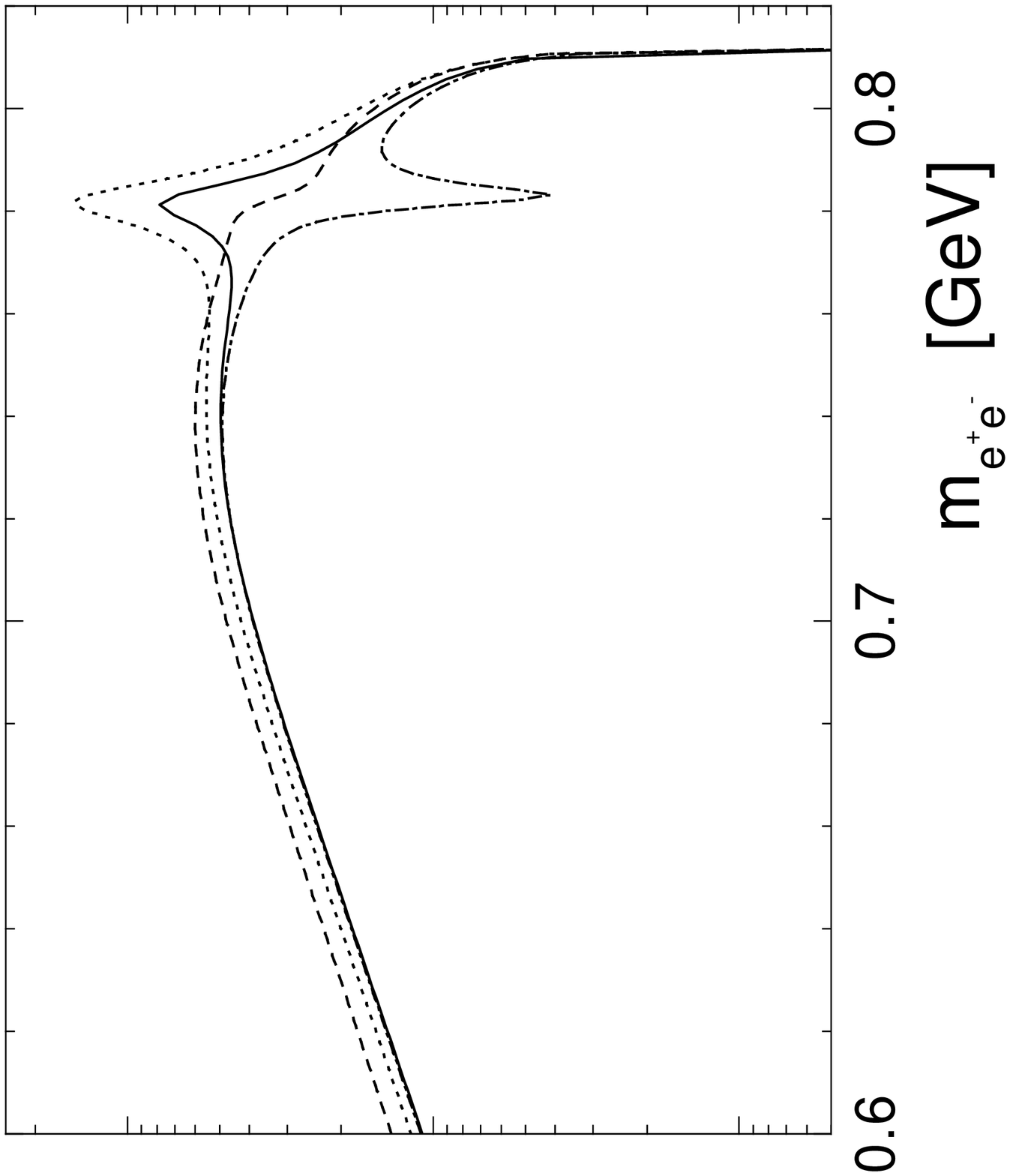, 
height=8cm}}}
\end{center}
\vspace{-8mm}
\caption[] 
{\label{picdilpnbor152}}
\end{figure}


\vspace{-1.5cm}

\begin{thebibliography}{239}

\itemsep0em

\bibitem{feuster}
T. Feuster and U. Mosel, Nucl. Phys. \textbf{A612}, (1997) 375.

\bibitem{iachello}
F. Iachello, A.D. Jackson and A. Lande, Phys. Lett. \textbf{B43}, (1973) 191. 

\bibitem{hohler}
G. H\"ohler, E. Pietarinen ad I. Sabba-Stefanescu, Nucl. Phys. \textbf{B114}, (1976) 505.

\bibitem{gari}
M. Gari and W. Kr\"umpelmann, Z. Phys. \textbf{A322}, (1985) 689; \newline
M. Gari and W. Kr\"umpelmann, Phys. Lett. \textbf{B173}, (1986) 10; \newline
M. Gari and W. Kr\"umpelmann, Phys. Lett. \textbf{B274}, (1992) 159. 

\bibitem{schafer}
M. Sch\"afer, H.C. D\"onges and U. Mosel, Phys. Rev. \textbf{C51}, (1995) 950.

\bibitem{mergell}
P. Mergell, U.-G. Mei\ss ner and D. Drechsel, Nucl. Phys. \textbf{A596}, (1996) 367. 

\bibitem{bijker}
R. Bijker and A. Leviatan, Contribution to CIPANP 97, available as \texttt{nucl-th/9706042} via WWW from \texttt{http://xxx.lanl.gov}. 

\bibitem{weidmann}
T. Weidmann, E.L. Bratkovskaya, W. Cassing and U. Mosel, submitted to Nucl. Phys. \textbf{A}, available as \texttt{nucl-th/9711004}.

\bibitem{gross}
F. Gross and D.O. Riska, Phys. Rev. \textbf{C36}, (1987) 1928.

\bibitem{nozawa}
S. Nozawa and T.-S.H. Lee, Nucl. Phys. \textbf{A513}, (1990) 511.

\bibitem{oconnell}
H.B. O'Connell et al., Phys. Lett. \textbf{B354}, (1995) 14, Prog. Part. Nucl. Phys. \textbf{39}, (1997) 201 and Nucl. Phys. \textbf{A623}, (1997) 559 and \texttt{hep-ph/9707253}; \newline
S. Gardner and H.B. O'Connell, \texttt{hep-ph/9707385}, submitted to Phys. Rev. \textbf{D}.

\bibitem{williams}
R.A. Williams, C.-R. Ji and S.R. Cotanch, Phys. Rev. \textbf{C46}, 1617.
\bibitem{manley}
D. M. Manley and E. M. Saleski, Phys. Rev. \textbf{D45}, (1992) 4002.

\bibitem{stoler}
P. Stoler, Phys. Rev. \textbf{D44}, (1991) 73 and Phys. Rep. \textbf{226}, (1993) 103.

\bibitem{goldberg}
H. Goldberg and Y. Srivastava, Phys. Rev. Lett. \textbf{22}, (1969) 749 
and Phys. Rev. Lett. \textbf{22}, (1969) 1332(E).

\bibitem{landsberg}
L. G. Landsberg, Phys. Rep. \textbf{128}, (1985) 300.

\bibitem{batzner}
K. B\"atzner et al., Phys. Lett. \textbf{B39}, (1972) 575.

\bibitem{lath79}
A. Latham et al., Nucl. Phys. \textbf{B156}, (1979) 58.

\bibitem{siddle}
R. Siddle et al., Nucl. Phys. \textbf{35}, (1971) 93.

\bibitem{shuttle}
W. J. Shuttleworth et al., Nucl. Phys. \textbf{B45}, (1972) 428.

\bibitem{alder}
J.-C. Alder et al., Nucl. Phys. \textbf{B105}, (1976) 253.

\bibitem{lath81}
A. Latham et al., Nucl. Phys. \textbf{B189}, (1981) 1.

\bibitem{evang}
E. Evangelides et al., Nucl. Phys. \textbf{B71}, (1974) 381.

\bibitem{morris78}
J. V. Morris et al., Phys. Lett. \textbf{B73}, (1978) 495.

\bibitem{wright}
J. Wright et al., Nucl. Phys. \textbf{B181}, (1981) 403.

\bibitem{aldpn}
J.-C. Alder et al., Nucl. Phys. \textbf{B99}, (1975) 1.

\end{thebibliography}
\end{document}